# Turning the LHC Ring into a New Physics Search Machine

Risto Orava[1, a)]

For the LHC Ring proto-collaboration

[1]*University of Helsinki, Helsinki Institute of Physics and CERN-EP, CH-1211 Geneva 23, Switzerland*

[a)]risto.orava@cern.ch

**Abstract.** The LHC Collider Ring is proposed to be turned into an ultimate automatic search engine for new physics in four consecutive phases: (1) Searches for heavy particles produced in Central Exclusive Process (CEP): $pp \to p + X + p$ based on the existing Beam Loss Monitoring (BLM) system of the LHC; (2) Feasibility study of using the LHC Ring as a gravitation wave antenna; (3) Extensions to the current BLM system to facilitate precise registration of the selected CEP proton exit points from the LHC beam vacuum chamber; (4) Integration of the BLM based event tagging system together with the trigger/data acquisition systems of the LHC experiments to facilitate an on-line automatic search machine for the physics of tomorrow.

## INTRODUCTION

### Goals of the Project

This proposal aims at configuring a new physics search facility based on existing instrumentation of the LHC ring and the LHC experiments. The approach presented here is novel, and uses the LHC Beam Loss Monitoring (BLM) and other LHC beam instrumentation devices for tagging the new physics event candidates in a model-independent way. The physics potential of the proposed facility is huge, and highly complementary to the present experimental installations at the LHC (ALICE, ATLAS/ALPHA, CMS/TOTEM, LHCb/MoEDAL experiments).

A few selected physics processes, based on Central Exclusive Production (CEP), Disoriented Chiral Condensates (DCCs), and gravitational waves are used as bench marks for the feasibility studies to be completed within the project. The CEP processes provide an ideal test ground for the proposed approach - here a pair of coincident final state protons, exiting the LHC beam vacuum chamber, are used to tag the event candidates. The fractional momenta of the final state protons are directly related to the invariant mass of the centrally produced system. The proposed approach [1,2] is independent of the particular decay modes of a centrally produced system, and substantially enhances the potential of observing new heavy particle states at the LHC. Performance of the customary Roman Pot technology is limited by the location of the pots, and the allowed transversal access to the beam.

The LHC Ring deforms due to low frequency gravitational wave background. It is of high importance to investigate further whether the LHC proton losses can be used for detecting faster transients (1ms to 10s frequency band) expected due to gravitational wave burst from different sources, such as binary black hole mergers [3] or other astrophysical phenomena.

The collaborators represent the key areas of this proposal: in accelerator physics and LHC instrumentation (S. Redaelli et al., CERN Beams Division), accelerator theory (Werner Herr, CERN Beams Division), theoretical high energy physics (Lucian Harland-Lang, University College, London, K. Huitu, Division of Particle Physics and Astrophysics, University of Helsinki; Valery Khoze, University of Durham University; M.G. Ryskin Petersburg Nuclear Physics Institute, Gatchina, St. Petersburg; V. Vento, University of Valencia and CSIC) and experimental high energy physics (A. De Roeck, CERN EP; M. Kalliokoski, CERN Beams Division; Beomkyu Kim, University of Jyväskylä; Jerry W. Lämsä, Iowa State University, Ames; C. Mesropian, Rockefeller University; Matti Mikael Mieskolainen, University of Helsinki; Toni Mäkelä, Aalto University, Espoo; Risto Orava, University of Helsinki,

Helsinki Institute of Physics and CERN; J. Pinfold, FRSC, Centre for Particle Physics Research, Physics Department, University of Alberta; Sampo Saarinen, University of Helsinki; M. Tasevsky, Institute of Physics of Academy of Sciences, Czech Republic)  and seismology (Pekka Heikkinen, Institute of Seismology, University of Helsinki).

The project has break-through potential both in high energy and gravitational wave physics. The basic infrastructure, the LHC Ring with its beam instrumentation and experiments, already exists, and only minor extensions are proposed for relatively inexpensive additional detectors and for facilitating triggering and automatic event selection. Preliminary analyses of the BLM signals (see the presentation by Matti Kalliokoski in this conference) validate the basic approach adopted by the authors, and include exiting candidate events in different physics categories listed below.

## SCANNING FOR NEW PHYSICS

## Central Exclusive Process - CEP

The Central Exclusive Production (CEP)  of particle state, $X$, is described by the following three processes:

$$pp(\gamma\gamma) \to p + (\gamma\gamma \to X) + p, \tag{1a}$$
$$pp(\gamma+gg) \to p + (\gamma g \to X) + p, \tag{1b}$$
$$pp(gg) \to p + (gg \to X) + p, \tag{1c}$$

where the + signs indicate rapidity gaps. The CEP sub-processes are facilitated by the photons (photon-photon interaction) (1a), photons and gluons ('photo-production' or 'photon-pomeron' interaction) (1b), and gluons ('diffractive' or 'double pomeron exchange'). In Figure 1.1, the corresponding Feynman diagrams for the processes (1a-c) are shown.

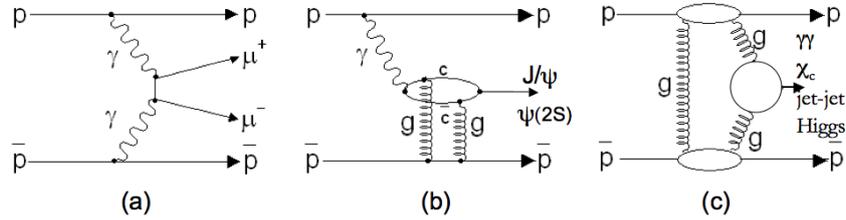

**Figure 1.1:** The Central Exclusive Production (CEP) processes facilitated by: (a) photon-photon sub interaction $pp(\gamma\gamma) \to p + (\gamma\gamma \to \mu^+\mu^-) + p$, (b) photon-gluon sub-*interaction* $pp(\gamma+gg) \to p + (\gamma g \to J/\Psi, \psi(2S)) + p$, and (c) gluon-gluon sub-interaction $pp(gg) \to p + (gg \to \chi_c, \text{jet-jet, Higgs}) + p$.

The respective cross sections for the processes (1a-c) are calculated as the convolutions of the effective luminosities $L(gg^{PP})$, $L(\gamma\gamma)$, or $L(\gamma g^{P})$ (Figure 1.2) and the square of the matrix element of the corresponding sub-process [4]. In the Central Exclusive Production (CEP) (1a-c), a number of advantageous properties exist compared to inclusive (or semi-inclusive) production: The mass and width of the centrally produced state, $X$, is correlated with the fractional (longitudinal) momentum losses, $\xi_{1,2} = 1 - p'_{z_{1,2}} / p_z$, of the final state protons ($p'_{z_{1,2}}$) and the initial beam proton ($p_z$), as:

$$M_X^2 \approx \xi_1 \xi_2 s, \tag{2}$$

where $s$ is the centre-of-mass energy squared. A measurement of the invariant mass of the decay products would be required to match the missing mass condition available by the measurement of the pair of final state proton fractional momentum losses. At higher central masses, $M_X \geq 200$ GeV, the photon-photon process (1a) dominates.

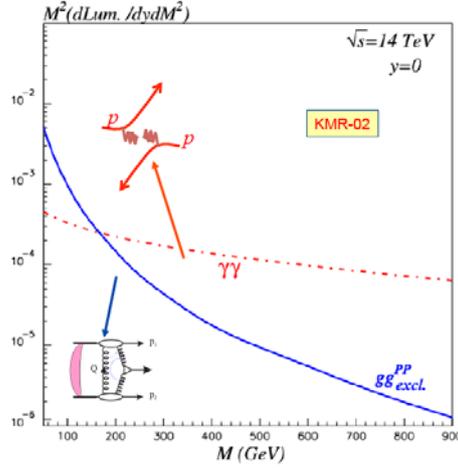

**Figure 1.2**: Gluon (solid line) and photon luminosities as a function of the central mass in Central Exclusive Production (CEP) [4].

The following example processes are considered: **(1) Magnetic monopolium**: Numerous experimental searches for magnetic monopoles have been carried out but all have met with failure. These experiments have led to a lower mass limit in the range from 350 − 500 GeV. A way out of this impasse is the above old idea of Dirac [5], namely, monopoles are not seen freely because they are confined by their strong magnetic forces forming a bound state called *monopolium* [6]. Here the CEP produced monopolium states [7] are searched for by registering the pair of final state protons exiting the beam vacuum chamber at the distance of some ±230 meters from IP8 (MoEDAL/LHCb experiment); **(2) Disoriented Chiral Condensates (DCCs)**: Bjorken et al. proposed [8] that high energy hadron collisions could form disoriented chiral condensates (DCCs): regions of pseudo-vacuum, where the chiral order parameter is misaligned from its vacuum orientation in isospin space. The formation of DCCs occurs when the highly Lorentz contracted nuclei collide and leave behind baryon free plasma in the central rapidity region where the chiral symmetry is subsequently restored. The plasma cools down through rapid hydrodynamic expansion via non-equilibrium chiral phase transition. Detection of such disoriented chiral condensate (DCC) would probe the chiral structure of the QCD vacuum and/or the chiral phase transition of strong interactions at high temperature. Here the LHCb experiment, with its beam-gas injection system provides a suitable framework for further studies [9]; **(3) $W^+W^-$ pairs and anomalous couplings**: Central exclusive production (CEP) of $W^+W^-$ ($Z^0Z^0$) pairs can be used both as a luminosity monitoring process [10] and as a process for studying basic physics questions beyond the Standard Model, such as anomalous vector boson couplings. In a recent publication the contribution of the $\gamma\gamma \to W^+W^-$ ($Z^0Z^0$) mechanism is compared to the gluon induced CEP process $gg \to W^+W^-$ [11]. The phase space integrated gluon induced CEP cross section is found to be considerably smaller (less than 1 fb), while the photon induced CEP is calculated to have a cross section of 115 fb. The photo-production process dominates at small four-momentum transfers for a wide range of $W^+W^-$ ($Z^0Z^0$) invariant masses, and allows efficient analyses of anomalous triple-boson (*WW/ZZ*) and quartic-boson (*WW/ZZ*) couplings together with tests of the models beyond the Standard Model. The $\gamma\gamma \to W^+W^-$ cross section peaks at $M_X \approx 200$ GeV yielding (in a symmetric case) a pair of protons exiting the beam vacuum chamber at ± 330 meters from the interaction point. All four LHC experiments have sufficient luminosity for studying the process; **(4) The Standard Model (SM) and BSM Higgs bosons**: The Higgs boson observations at the LHC are almost exclusively based on the $\gamma\gamma$ decay mode [12]. Measurement of the Higgs boson production in Central Exclusive Process (CEP) was first analysed by some of the authors, and the process $pp \to p + (gg \to h^0) + p$ (1c) provides important complementary information concerning the spin-parity state of Higgs since $J^{PC} = 0^{++}$ state is strongly favoured in CEP. By tagging the Higgs event candidates independently of the Higgs decay products enables detailed analysis of the production mechanism and Higgs couplings. A measurement of the azimuthal angle between the final state protons can be used to discriminate between different Higgs production scenarios [13]. The Standard Model Higgs boson, when produced in the CEP process (1c), has a proton pair exiting the LHC beam vacuum chamber at a distance of ± 427 meters from the IP. ATLAS, CMS and LHCb experiments are here relevant counterparts due to their sufficiently high integrated luminosities; **(5) Gluon jets**: The jets produced in the CEP process: $pp \to p + X + p$, are practically always gluon jets, the purity vs. light quark jets is of the order of $10^3$ [14]. The CEP jet final states can therefore be used for investigating gluon jets in detail, thereby turning the LHC into a gluon factory. Observables such as jet multiplicities, flavour and baryon number suppression can be accurately measured and compared to the inclusive minimum bias interactions. The uniquely pure gluon jet final states are available via the CEP process in all LHC experiments. The ALICE experiment, with its relatively low pile-up rates, has access to CEP produced gluon jets in the mass range of some 40-60 GeV; these events are tagged by the final state proton pairs exiting at ± 400 meters and beyond from IP2; **(6) Gravitational waves**: Fast transients caused by the gravitational waves have been recently observed by the Laser Interferometer Gravitational-Wave Observatory (LIGO) [3] in the frequency range of 35 to 250 Hz as predicted by Einstein's general theory of relativity. Gravity will influence the storage rings used in particle physics accelerator systems, as shown earlier for the Large Electron Positron Collider (LEP) [15] at CERN and for the SPRing-8 in Japan [16]. According to [17], the change $\Delta C$ of the reference value of the machine circumference changes due to the gravitational interaction induced by the tidal and seasonal forces. An overall change of the storage ring circumference can be expressed as

$$\Delta C / \Delta t = (\Delta C / \Delta t)_m + (\Delta C / \Delta t)_s + (\Delta C / \Delta t)_{gw} , \qquad (3)$$

where the two first terms are caused by the tidal ($m$) and seasonal ($s$) forces, and the third term by the cosmic very low frequency gravitational wave background ($gw$). The net change in the machine circumference (Equation (3)) has been found to be about $2 \times 10^{-4}$ m/yr [18], and cannot be explained by the tidal and seasonal forces induced by the Moon and Sun alone. In this analysis, the transients induced by gravitational waves in the frequency ranges between $10^{-3}$ to 10 seconds are analysed in terms of the signal correlations measured by the LHC BLM system around the ring. In a preliminary analysis of the BLM signals, wave forms expected from the recent nuclear tests in North Korea, and from recent known earth quakes are clearly seen by the LHC BLM system, and represent the background to be investigated in detail for calibrating and evaluating the capabilities of the LHC machine to register gravitational waves [19].

## CEP PROTONS EXITING THE LHC RING

By tracing CEP protons of different $\xi$-values through the LHC accelerator lattice [1], a relation between the CEP proton exit points and the $\xi$-values of the final state protons is established. For the background studies of this proposal, both $\xi$ and the transverse momentum of $p_T$ of the final state protons are considered in mapping out the exit points around the LHC ring.

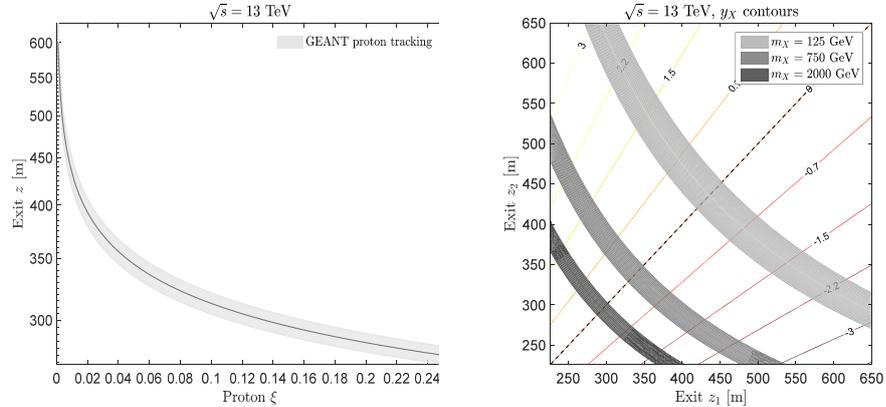

**Figure 1.3: Left panel**: The proton exit point, $z$, in CEP: $pp \rightarrow p + X + p$, as a function of the fractional momentum loss, $\xi$ (solid line). The exit points of the leading protons out from the beam vacuum chamber are given in meters from IP5, the shaded band reflects smearing in proton transverse momentum. **Right panel**: The proton exit point combinations in CEP: $pp \rightarrow p + X + p$, as a function of the central mass, $M_X$ (the grey bands). The exit points of the leading protons out from the beam vacuum chamber are given in meters from the Interaction Point 5 (IP5), the symmetric cases ($\xi_1 \approx \xi_2$) have $z_1 \approx z_2$ (dashed diagonal line). The rapidity span of the centrally produced decay products scales as $\Delta y \propto \ln(M_X^2)$ (solid lines with the rapidity scale), rapidity of the centrally produced state is given as $y_X = 0.5\ln(\xi_1/\xi_2)$ [1].

In Figure 1.3 (left panel), the proton exit points, shown as a function of their fractional momentum loss, $\xi$, are produced by the proton tracing codes. Through Equation (2), the measured proton exit locations can then be used for an $M_X$ mass scan of the centrally produced systems (Figure 1.3, right panel). The band widths reflect smearing in proton transverse momentum, $p_T$. The following steps are taken in tagging the CEP event candidates for each IP (IP1/ATLAS, IP2/ALICE, IP5/CMS, and IP8/LHCb): (i) The candidate CEP events are scanned by locating pairs of coincident proton exits on the opposite sides of the interaction point (IP) in question (Figure 1.3, right panel), (ii) The tagged events are correlated with the LHC Beam Cross Overs (BCOs) within the time window for the chosen IP, (iii) The tagged LHC BCOs are analysed as candidates for the CEP events with central masses, $M_X$, corresponding to a registered pair of exit points (Figure 1.3, right panel).